\documentclass[aps,pra,10pt,showpacs,onecolumn,superscriptaddress,twocolumn,longbibliography]{revtex4-1}

\usepackage{graphicx,caption}
\usepackage{booktabs}
\usepackage[usenames]{color}
\usepackage{amssymb,amsmath}
\usepackage{threeparttable}
\usepackage{siunitx}
\usepackage{xcolor}
\graphicspath{{}}
\usepackage{setspace} 
\usepackage{float} 
\usepackage{tabularx}
\usepackage[linkcolor=blue,citecolor=blue,colorlinks=true,urlcolor=blue]{hyperref}

\newcommand{\be}[1]{\begin{equation}\label{#1}}
\newcommand{\ee}{\end{equation}}
\usepackage{amsmath,esint}
\usepackage{comment}
\usepackage{lipsum}
\usepackage[toc]{appendix}
\begin{document}

\title{Extracting  inter-nuclear distances in  the oxygen molecule interacting with an XFEL-pulse: a fundamental system for understanding  Coulomb explosion imaging   }

\author{Zixu Wang}
\affiliation{Department of Physics and Astronomy, University College London, Gower Street, London WC1E 6BT, United Kingdom}
\author{
Georgios Petros Katsoulis}
\affiliation{Department of Physics and Astronomy, University College London, Gower Street, London WC1E 6BT, United Kingdom}
\author{
Giorgio Visentin}
\affiliation{Department of Physics and Astronomy, University College London, Gower Street, London WC1E 6BT, United Kingdom}
\author{
Till Janke} 
\affiliation{Institute for Nuclear Physics, Johann Wolfgang Goethe-University Frankfurt, Max-von-Laue-Str. 1, 60438 Frankfurt, Germany}
\author{
Thomas Pfeifer}
\affiliation{Max Planck Institute for Nuclear Physics, Saupfercheckweg 1, 69117 Heidelberg, Germany}
\author{
Agapi Emmanouilidou}
\affiliation{Department of Physics and Astronomy, University College London, Gower Street, London WC1E 6BT, United Kingdom}

\begin{abstract}
We investigate the interaction of O$_{2}$ with an X-ray Free Electron laser (XFEL) pulse of short duration. We consider a photon energy of 570 eV, which allows for the formation of  molecular states with up to two core holes.    We compute the sum of the final kinetic energies, i.e. the kinetic energy release  (KER) of the  atomic fragments for different fragmentation channels. We demonstrate that  with the knowledge of the potential energy curves the equilibrium inter-nuclear distance of O$_{2}$ is best extracted from the O$^{+}$+O$^{+}$ channel and not from higher-charged channels such as O$^{2+}$+O$^{2+}$. This challenges  our current understanding of Coulomb explosion imaging, where from just applying quasi-classical dynamics  one expects that the equilibrium inter-nuclear distance of molecules is best extracted from  higher-charged fragmentation channels. On the other hand, we show that the  KER of the O$^{+}$+O$^{+}$  channel is inconsistent with the one obtained by just calculating   the Coulomb repulsion of the atomic fragments resulting from molecular dissociation.
\end{abstract}

\maketitle

\section{\label{sec:level1}Introduction}
Capturing structural dynamics as a function of time  provides fundamental insights into chemical reactions. Coulomb explosion imaging (CEI)  correlates  molecular geometry at the time when Coulomb explosion occurs with the final momenta of the atomic fragments.   It is widely accepted that CEI captures the initial  molecular structure by ionizing  a molecule fast into a high charge state, resulting   to a rapid fragmentation  due to Coulomb repulsion \cite{ref7,ref8,ref9,ref10,ref11}.


X-ray free electron lasers (XFELs) enable inner shell photoionization followed by Auger-Meitner decay allowing for populating high charge states and leading to quick Coulomb explosion  \cite{ref10,ref13}. Hence, XFELs are a powerful tool for enabling  CEI \cite{ref14,ref15,ref16}, since they allow to populate molecular states  whose potential surfaces are mostly influenced by the Coulomb repulsion, baring almost no features of chemical bonding. CEI has been employed to investigate a wide range of photo-physics and photo-chemistry processes, such as photodissociation \cite{ref11,ref19,ref25}, roaming \cite{ref9}, conical intersections \cite{ref26} and to enable characterization of isomers \cite{ref23,ref27,ref28}. Hence, CEI significantly contributes to increasing our understanding of chemical reactions and molecular dynamics.

 An assumption involved in XFEL-driven CEI being able to image the initial molecular structure is that XFELs achieve a quick transition to highly charged molecular states. Moreover, it is also assumed that in these highly charged states the atoms Coulomb explode fast with minimal influence from chemical bonding. These assumptions allow to complement CEI experiments by  simply employing  classical  equations of motion that only incorporate the Coulomb repulsion of the atomic ions in the molecule, see for instance Ref.\cite{ref7}.

 Employing ab-initio theoretical techniques 
that fully address molecular structure and chemical bonding is highly challenging. This is more so for large molecules, at the center of interest in CEI experiments in recent years. However, it has been pointed out that for lower charged molecular states ab-initio theoretical studies are needed. Indeed, in these states, chemical bonding strongly influences the potential surfaces, with Coulomb repulsion being just one  of the forces present \cite{ref15,ref17,ref18,ref25,ref30}.

Here, in the context of a diatomic molecule, O$_{2}$, we shed more light into imaging the initial molecular structure using the final momenta of the atomic fragments, and in particular the sum of the final kinetic energies of the two atomic fragments, i.e. the kinetic energy release (KER). Contrary to wide expectation, we show that for O$_{2}$, it is fragmentation via the lower charged molecular states that allows for accurate extraction of the initial (equilibrium) inter-nuclear distance of the ground state. We demonstrate that this extraction from the KER of the two O fragments    
is better achieved from the lower charged molecular states and, hence, is not related primarily to Coulomb explosion. However, this extraction is achieved using the potential energy curves of the molecule. We show that extracting the equilibrium inter-nuclear distance  from  dissociating higher-charged states of O$_{2}$,  where the dominant force is Coulomb repulsion among the atomic fragments, is not as accurate as from the lower-charged states.

We employ a fully ab-initio theoretical framework to describe electronic structure and compute molecular continuum states of escaping electrons \cite{Banks2017N2FEL}. Moreover, we employ the classical two-body equations of motion to obtain the momenta of the atomic fragments \cite{Mountney2025}. However, the potential employed in these classical equations is obtained with fully ab-initio advanced quantum chemistry techniques \cite{Mountney2025}. The photon energy of the XFEL pulse we employ is 570 eV which allows for two core electrons to escape, each from one of the two core molecular orbitals.  We consider fragmentation channels where the  total charge of the atomic fragments is up to four. We show that our findings are quite robust and hold for a range of pulse durations and photon energies.







\section{THEORETICAL MODEL}\label{Sec::THEORETICAL_MODEL}

  In Ref.~\cite{Mountney2025}, we developed a hybrid quantum-classical model  to describe the interaction of extreme-ultraviolet (XUV) laser pulses with diatomic molecules. Here, we expand this hybrid model  to account for  the interaction of diatomics with X-ray pulses. We consider O$_2$ interacting with XFEL pulses with photon energy  sufficient to ionize a core electron  from each of the two core-orbitals, i.e. a total of two core electrons. In this model, the electron dynamics is treated quantum mechanically. The nuclear dynamics is treated using classical equations of motion, while employing quantum-mechanical potential-energy curves (PECs). We compute the PECs for molecular charge states up to O$_2^{5+}$ using advanced quantum-chemistry methods, see Sec.~\ref{Sec::PECs_calculation}. The Born-Oppenheimer approximation is adopted to treat both the electronic and nuclear degrees of freedom. We combine in a Monte-Carlo, i.e. stochastic, method, the quantum-mechanical treatment of the electron motion with the hybrid description of nuclear motion.   Concerning core-hole states, we account for electronic states containing up to two core holes located at different molecular orbitals.  In Sec.~\ref{Sec::Electronic_process}, we briefly describe how we  obtain quantum mechanically the single-photon ionization cross-sections and Auger-Meitner rates for all electronic  transitions energetically allowed given the photon energy of the laser pulse. In Sec.~\ref{Sec::Molecular_dissociation_process}, we outline how we determine  the inter-nuclear distance where we transition from a molecule  to two atomic fragments interacting with the XFEL-pulse. The fragments result from dissociation of the molecular states.  In Sec.~\ref{Sec::Nuclear_propagation}, for completeness, we briefly describe the two-body system classical equations of motion, which are used to time propagate  the nuclei.   In Sec.~\ref{Sec::Sampling}, we describe sampling the initial position and momentum of the nuclei 
at the start of the propagation using the Wigner distribution of the ground state of the harmonic oscillator. Incorporating all the above elements in a Monte Carlo method (Sec.~\ref{Sec::Monte_Carlo_simulation}), we simulate the interaction of O$_2$ with an XFEL-pulse. 

\subsection{Computation of PECs}\label{Sec::PECs_calculation}
For 570 eV photon energy, we compute and account for all 927 energetically accessible states of molecular oxygen up to O$_2^{5+}$, with up to two core holes, one from each core orbital.  These states are pertinent for photon energies roughly up to 603 eV at all inter-nuclear distances.  Since we compute molecular states up to O$_2^{5+}$,  we consider fragmentation channels where the atomic-ion fragments have total  charge up to four. Each state of O$_2$ is denoted by its electronic configuration $\mathrm{(1\sigma_g^{a}, 1\sigma_u^{b}, 2\sigma_g^{c}, 2\sigma_u^{d}, 3\sigma_g^{e} 1\pi_{ux}^{f}, 1\pi_{uy}^{g}, 1\pi_{gy}^{h}, 1\pi_{gx}^{i})}$, where a, b, c, d, e, f, g, h, and i are the occupation numbers of each molecular orbital, with occupancy 0, 1, or 2. The core orbitals are $1\sigma_g$ and $1\sigma_u$, $2\sigma_g$ is the inner valence orbital, while the remaining orbitals  are valence ones~\cite{Mountney2025}. 

	First, we use the Hartree-Fock (HF) method to compute the bound orbitals of the ground state of O$_2$ as a function of the inter-nuclear distance. These orbitals are subsequently used as input for the calculation of the potential-energy curves with the complete active space self-consistent field (CASSCF) method. We use a complete active space of ten orbitals, composed of nine occupied orbitals and one virtual $3\sigma_u$ orbital, in combination with the augmented Dunning correlation-consistent quadruple-$\zeta$ basis set (aug-cc-pVQZ)~\cite{Mountney2025, Hadjipittas2023}. We compare the PECs from the available literature~\cite{Hikosaka2003,Bao2008,Yagishita1994} with the PECs we obtain in this work and find good agreement. Hence, we generate with sufficient accuracy and robustness the database of PECs for the Monte-Carlo simulations of the X-ray-driven molecular dynamics. The  PECs are computed from \(1.51~\mathrm{a.u.}\) (\(0.8~\text{\AA}\)) to \(7.75~\mathrm{a.u.}\) (\(4.1~\text{\AA}\)) in steps of \(0.19~\mathrm{a.u.}\) (\(0.1~\text{\AA}\)), since the distance of  \(7.75~\mathrm{a.u.}\) (\(4.1~\text{\AA}\)) is sufficient for the molecular states with charge 3 and higher to be determined by Coulomb repulsion. Molecular states with charge 1 converge to their final values at distances less than \(7.75~\mathrm{a.u.}\) However, for molecular charge states with charge 2, mostly those with core holes, the chemical bonding affects the PECs at distances much larger than \(7.75~\mathrm{a.u.}\) and as a result we compute these potential energy curves up to significantly larger distances.  All the computations of the PECs are implemented within the framework of the quantum-chemistry package MOLPRO~\cite{Werner2012Molpro}.
	Two issues are  addressed during the computation of the 927 PECs. Namely, in some cases,  the molecular state of interest is not the lowest-energy state of a given symmetry. As a result, variational collapse occurs~\cite{Besley2009, Bhattacharya2021}, with MOLPRO computing the PEC for the lowest-energy state within that symmetry rather than the desired state. To address this, we employ state-averaging CASSCF (SA-CASSCF) to compute the PECs, following the strategy employed in Refs~\cite{Bhattacharya2021, Hadjipittas2023}. 
	 The equilibrium inter-nuclear distance of O${_2}$ obtained with CASSCF is $r_e = 2.296~\text{a.u.}  (1.215~\text{\AA})$, in agreement with previously reported values~\cite{Bytautas2010,Weast1985}. Another issue is the computation of PECs for molecular states with at least one inner-shell hole and any other combination of electrons missing from valence orbitals. For this computation, we employ a two-step (TS) optimization technique, either the TS-CASSCF or the SA-TS-CASSCF approach~\cite{Hadjipittas2023,Bhattacharya2021}. The former approach is used when the target state is the lowest-energy state of the given symmetry, whereas the latter is used otherwise. In these two approaches, in the first step, the valence orbitals are optimized while the core orbitals $1\sigma_g$ and $1\sigma_u$ are kept frozen. In the second step, the core orbitals are optimized while the valence orbitals are kept frozen \cite{Rocha2011,Carravetta2013,Bhattacharya2021}. Similar techniques are employed to obtain the energies of the energetically accessible states of the atomic oxygen fragments. For the small pulse duration we consider in the current work and given that we consider molecular states up to O$_2^{5+}$, it is sufficient to consider atomic ions with charge up to 6,  accounting for 231  ion states of O. 

\subsection{Single photonionization cross sections and Auger-Meitner rates}\label{Sec::Electronic_process}

Here, the photon energy considered is sufficient to create up to two core holes due to single photoionization, one on each of the core orbitals of O$_2$. As a result, Auger-Meitner processes play a significant role in the interaction of XFEL-pulse with O$_2$. In the current work, we consider two main electronic processes, namely single photoionization (SPI) and Auger-Meitner decay (AM), for both the molecule and the atomic fragments interacting with the XFEL-pulse. To compute molecular SPI cross sections and AM decay rates, we adopt the \textit{ab initio} framework of Refs.~\cite{Banks2017N2FEL, Demekhin2011SCM} for the calculation of continuum wave functions. The continuum orbitals are computed using the Hartree–Fock (HF) equations \cite{Bransden1983AtomsMolecules}. The bound orbitals of the initial molecular electronic state are obtained using the Hartree–Fock method  in the framework of MOLPRO~\cite{Werner2012Molpro}. Following the method described in Refs.~\cite{Banks2017N2FEL, Demekhin2011SCM}, we solve a system of  Hartree-Fock equations to obtain the continuum orbitals.  Both the bound and continuum wave functions are expressed in terms of the single-center expansion (SCE) scheme~\cite{Banks2017N2FEL, Demekhin2011SCM}.
\begin{align}
\psi(\mathbf{r}) = \sum_{lm}\frac{P_{lm}(r)Y_{lm}(\theta,\phi)}{r},
\label{eq:sce_wavefunction}
\end{align}
where \(l\) and \(m\) denote the angular and magnetic quantum numbers, respectively, \(Y_{lm}(\theta,\phi)\) is a spherical harmonic, and \(P_{lm}(r)\) is the radial single-center expansion coefficient for the partial-wave channel \((l,m)\).

	The SPI cross section for an electron transitioning from an initial bound orbital \(\psi_i\) to a final continuum orbital \(\psi_{\epsilon}\) is given by Refs.~\cite{Sakurai1994,Banks2020CO}
\begin{align}\label{eq:photoionization_cross_section}
\begin{split}
\sigma_{i\to\epsilon}
= \frac{4}{3}\alpha\pi^2\omega N_i
\sum_{M=-1,0,1}\left|D_{i\epsilon}^{M}\right|^2,
\end{split}
\end{align}
where \(\alpha\) is the fine-structure constant, \(\omega\) is the photon energy, \(N_i\) is the occupation number of orbital \(i\) before the electron escapes to the continuum, and \(M\) denotes the polarization of the photon, which takes the values \(-1\), \(0\), or \(1\). By inserting Eq.~\eqref{eq:sce_wavefunction} in the expression of the dipole matrix element \(D_{i\epsilon}^{M}\), we obtain \(D_{i\epsilon}^{M}\) in the length gauge as follows
\begin{widetext}
\begin{align}
\label{eq:dipole_matrix_element}
D_{i\varepsilon}^{M}
&= \left\langle \phi_{\varepsilon} \middle| d_{M}^{L} \middle| \phi_i \right\rangle  \notag\\
&= \int \phi_{\varepsilon}^{*}(\mathbf{r})
\sqrt{\frac{4\pi}{3}}\, r Y_{1M}(\theta,\phi)\,
\phi_i(\mathbf{r})\, d\mathbf{r} \notag\\
&= \sqrt{\frac{4\pi}{3}}
\sum_{lm,l'm'}
\int_{0}^{\infty} dr\,
P_{\varepsilon,l'm'}^{*}(r)\, r\, P_{i,lm}(r)
\int d\Omega\,
Y_{l'm'}^{*}(\theta,\phi)
Y_{lm}(\theta,\phi)
Y_{1M}(\theta,\phi) \notag\\
&= \sum_{lm,l'm'}
(-1)^{m'}
\sqrt{(2l+1)(2l'+1)}
\begin{pmatrix}
l' & l & 1 \\
0 & 0 & 0
\end{pmatrix}
\begin{pmatrix}
l' & l & 1 \\
-m' & m & M
\end{pmatrix}
\int_{0}^{\infty} dr\,
P_{\varepsilon,l'm'}^{*}(r)\, r\, P_{i,lm}(r).
\end{align}
\end{widetext}
	Homonuclear molecules have cylindrical symmetry about the inter-nuclear axis and as a result, the magnetic quantum number \(m\) is a good quantum number. Hence, the summation in Eq.~\eqref{eq:dipole_matrix_element} is only over \(l\) values while m has a specific value dictated by the m number of the initial molecular orbital relevant to the SPI or AM process under consideration.  Moreover, inversion symmetry for homonuclear molecules results in molecular orbitals having gerade or ungerade parity, which restricts the contributing partial waves to either even or odd values of \(l\)~\cite{Banks2019Thesis, Mountney2024Thesis}.

	The total AM rates, denoted by \(\Gamma\), are obtained using Fermi's golden rule~\cite{Pauli2000WaveMechanics} as follows
\begin{align}\label{eq:auger_rate_general}
\begin{split}
\Gamma
= \overline{\sum} 2\pi |\mathcal{M}|^2
\equiv
\overline{\sum} 2\pi \left|\left\langle \Psi_{\mathrm{fin}} \middle| H_I \middle| \Psi_{\mathrm{init}} \right\rangle\right|^2,
\end{split}
\end{align}
where $\overline{\sum}$ denotes a sum over the final states and an average over the initial states. The wave functions of the initial and final molecular states are denoted by \(\Psi_{\mathrm{init}}\) and \(\Psi_{\mathrm{fin}}\), respectively, while \(H_I\) represents the interaction Hamiltonian. The interaction Hamiltonian \(H_I\) corresponds to the electron-electron Coulomb interaction. In the second-quantization formalism~\cite{Manne1985,Inhester2013,Banks2017N2FEL}, we obtain
\begin{align}\label{eq:H_I}
\begin{split}
H_I^{\mathrm{ee}}
= \frac{1}{2}\sum_{\alpha\beta\gamma\delta}
c_{\alpha}^{\dagger} c_{\beta}^{\dagger} c_{\gamma} c_{\delta}
\left\langle \alpha\beta \middle| \frac{1}{r_{12}} \middle| \gamma\delta \right\rangle,
\end{split}
\end{align}
where \(c_{\gamma}\) is the annihilation operator associated with the one-electron spin-orbital \(|\gamma\rangle\), and \(c_{\alpha}^{\dagger}\) is the corresponding creation operator for the one-electron spin-orbital \(|\alpha\rangle\).

To obtain the group AM rate, without resolving the spin of the electrons, we insert Eq.~\eqref{eq:H_I} in Eq.~\eqref{eq:auger_rate_general} and express \(\Psi_{\mathrm{init}}\) and \(\Psi_{\mathrm{fin}}\) with single Slater determinants. 
We obtain in the \(m_a, m_b, S, M_S\) scheme the partial AM rate for a transition between two electronic configurations as follows~\cite{Mountney2024Thesis}
\begin{align}\label{eq:auger_rate_abs}
\begin{split}
\Gamma_{a,b\to s}
=\\
\sum_{S M_S}
\pi N_h N_{ab}
\sum_{\zeta}
\left| \left\langle \zeta s \middle| \frac{1}{r_{12}} \middle| ba \right\rangle + (-1)^S \left\langle \zeta s \middle| \frac{1}{r_{12}} \middle| ab \right\rangle \right|^2,
\end{split}
\end{align}
where \(N_h\) is the number of holes in the \(s\) orbital. The indices \(a\) and \(b\) denote valence orbitals and \(\zeta\) labels the different degenerate channels of the continuum orbital. \(S\) and \(M_S\) are the total spin and its projection of the two valence electrons involved in the AM transition, and \(N_{ab}\) is the weighting factor related to the occupation numbers of the valence orbitals a and b 
\begin{equation}
\label{eq:Nab}
N_{ab}
=
\begin{cases}
\dfrac{N_a N_b}{2 \times 2}, 
& \text{for electrons in different orbitals}, \\[1.2em]
\dfrac{N_a(N_a-1)}{2}, 
& \text{for electrons in the same orbital}.
\end{cases}
\end{equation}
 Inserting the single-center expansion of both the continuum and bound orbitals  in Eq.~\eqref{eq:auger_rate_abs}, we obtain the AM rate as follows

\begin{widetext}
\begin{equation}
\label{eq:AugerRates}
\begin{aligned}
&
\left\langle \zeta s \middle| \frac{1}{r_{12}} \middle| ba \right\rangle
+(-1)^S
\left\langle \zeta s \middle| \frac{1}{r_{12}} \middle| ab \right\rangle
\\
&\qquad =
\sum_{\substack{k l_{\zeta} l_s\\ l_b l_a}}
\sum_{q=-k}^{k}
\int dr_1 \int dr_2\,
P_{l_{\zeta}m_{\zeta}}^{\zeta *}(r_1)
P_{l_s m_s}^{s *}(r_2)
\frac{r_{<}^{k}}{r_{>}^{k+1}}
P_{l_b m_b}^{b}(r_1)
P_{l_a m_a}^{a}(r_2)
\\
&\qquad\quad\times
(-1)^{m_s}
\sqrt{(2l_s+1)(2l_a+1)}
\begin{pmatrix}
l_s & k & l_a \\
0 & 0 & 0
\end{pmatrix}
\begin{pmatrix}
l_s & k & l_a \\
-m_s & q & m_a
\end{pmatrix}
(-1)^{q+m_{\zeta}}
\sqrt{(2l_{\zeta}+1)(2l_b+1)}
\\
&\qquad\quad\times
\begin{pmatrix}
k & l_{\zeta} & l_b \\
0 & 0 & 0
\end{pmatrix}
\begin{pmatrix}
k & l_{\zeta} & l_b \\
-q & -m_{\zeta} & m_b
\end{pmatrix}
\\[1ex]
&\qquad
+(-1)^S
\sum_{\substack{k l_{\zeta} l_s\\ l_b l_a}}
\sum_{q=-k}^{k}
\int dr_1 \int dr_2\,
P_{l_{\zeta}m_{\zeta}}^{\zeta *}(r_1)
P_{l_s m_s}^{s *}(r_2)
\frac{r_{<}^{k}}{r_{>}^{k+1}}
P_{l_b m_b}^{b}(r_1)
P_{l_a m_a}^{a}(r_2)
\\
&\qquad\quad\times
(-1)^{m_s}
\sqrt{(2l_s+1)(2l_b+1)}
\begin{pmatrix}
l_s & k & l_b \\
0 & 0 & 0
\end{pmatrix}
\begin{pmatrix}
l_s & k & l_b \\
-m_s & q & m_b
\end{pmatrix}
(-1)^{q+m_{\zeta}}
\sqrt{(2l_{\zeta}+1)(2l_a+1)}
\\
&\qquad\quad\times
\begin{pmatrix}
k & l_{\zeta} & l_a \\
0 & 0 & 0
\end{pmatrix}
\begin{pmatrix}
k & l_{\zeta} & l_a \\
-q & -m_{\zeta} & m_a
\end{pmatrix}.
\end{aligned}
\end{equation}
\end{widetext}
A detailed derivation of the AM rate is given in Ref.~\cite{Mountney2024Thesis}.

The truncated \(l\) values  in Eq.~\eqref{eq:dipole_matrix_element} and Eq.~\eqref{eq:AugerRates} are determined so that convergence is achieved  for SPI cross sections and AM decay rates. We find the truncated $l$ value to be 40 for the bound molecular orbitals,  and 90 and 35, for the continuum  orbitals  for the  SPI cross sections and AM rates, respectively.

	 We compute the SPI cross sections and AM rates  for all energetically allowed transitions at 570 eV photon energy. The calculations are performed for internuclear distances ranging from \(1.51~\mathrm{a.u.}\) (\(0.8~\text{\AA}\)) to \(3.97~\mathrm{a.u.}\) (\(2.1~\text{\AA}\)) in steps of \(0.19~\mathrm{a.u.}\) (\(0.1~\text{\AA}\)), to account for the rapid variation of the PECs around the equilibrium distance, and from \(3.97~\mathrm{a.u.}\) (\(2.1~\text{\AA}\)) to \(7.75~\mathrm{a.u.}\) (\(4.1~\text{\AA}\)) in steps of \(0.38~\mathrm{a.u.}\) (\(0.2~\text{\AA}\)). At each distance, we compute   2771 cross sections for the energetically allowed  single photon transitions  and 5291 rates for the energetically allowed AM transitions. Interpolating the   SPI cross sections and AM rates  with the internuclear distance, we then  incorporate them  in  the Monte-Carlo simulations described later in this work. 
	 
	For the atomic fragments resulting from the dissociation of the molecular state, we compute the continuum wave functions of the different electronic configurations using the well-established Hartree-Fock-Slater method~\cite{Slater1951,Riley1975,Herman1963}, as we have done in  previous work~\cite{Wallis2014ArgonAugerSpectra}. We then obtain the atomic SPI cross sections and AM rates following the method outlined in Refs.~\cite{Wallis2014ArgonAugerSpectra,Bhalla1973NeonAuger}. For 570 eV photon energy, we compute 625  cross sections for the energetically allowed single photon transitions and  638 rates for the energetically allowed AM  transitions.

\subsection{Transition from  molecular to atomic-fragment dynamics in the Monte-Carlo simulation }\label{Sec::Molecular_dissociation_process}
Computation of PECs, as well as of molecular SPI cross sections and AM rates is significantly more demanding than the computation of SPI cross sections and AM rates for atomic fragments. Moreover, the computation of molecular SPI and AM rates becomes less accurate at large inter-nuclear distances, since we employ HF orbitals to compute these rates which fail to accurately describe PECs at large inter-nuclear distances. Hence, in the Monte-Carlo simulation, we transition at a certain inter-nuclear distance from the dissociating molecular state interacting with the XFEL-pulse to the resulting two atomic fragments interacting with the XFEL-pulse. To find this inter-nuclear distance, 
 we  identify the distance where the PECs follow the Coulomb repulsion curves between the two atomic ion fragments. We find this inter-nuclear distance to be  \(4.1~\text{\AA}\) for most PECs. Namely, the PECs for molecular states with charge 3 and higher follow the Coulomb repulsion curve at     \(4.1~\text{\AA}\), while 
  the molecular states with charge 1 converge to their asymptotic values before the inter-nuclear distance of \(4.1~\text{\AA} \). However, several molecular states with charge 2, and mostly those with core holes, follow  the Coulomb repulsion curves at significantly larger distances. That is, chemical bonding plays a significant role even at  large distances for several molecular ion states  with charge 2.  Hence, we compute several molecular PECs with charge 2 up to much large distances than  \(4.1~\text{\AA} \).   As a result, in the Monte-Carlo simulation, we transition from a molecule to the two atomic fragments at  \(4.1~\text{\AA}\) for all molecular states. However,   for the molecular states with charge 2 that their PECs do not agree with the Coulomb repulsion curves beyond \(4.1~\text{\AA} \), we propagate the atomic fragments on the molecular curves up to \(20~\text{\AA} \). For these latter states, once the distance of \(20~\text{\AA} \) is reached in the Monte-Carlo simulation, only then do we consider the potential energy equal to the  Coulomb repulsion of the atomic fragments.

To identify the atomic fragments resulting from molecular dissociation, we first compute the energy of each molecular ion state at an asymptotic inter-nuclear distance. We use this energy as the dissociation limit of each molecular ion state. Next, for each molecular state, we generate all possible combinations, in terms of electronic configurations, of the two atomic ion fragments with the same total charge as the molecular ion. We select the states of the atomic ions that have total energy closest to the corresponding molecular dissociation limit. These  atomic fragments replace the molecular state  in the Monte-Carlo simulation, as discussed above.

\subsection{Propagation of the nuclei}
\label{Sec::Nuclear_propagation}
In  the classical two-body description, 
the position and velocity of the nuclei are given by 
\begin{equation}
\label{eq:Velocities}
\begin{aligned}
\boldsymbol{x}_1 &= \frac{\mu}{m_1}\boldsymbol{r} 
\; \Rightarrow\; \boldsymbol{v}_1 = \dot{\boldsymbol{x}}_1 = \frac{\mu}{m_1}\dot{\boldsymbol{r}}, \\
\boldsymbol{x}_2 &= -\frac{\mu}{m_2}\boldsymbol{r} 
\;\Rightarrow\; \boldsymbol{v}_2 = \dot{\boldsymbol{x}}_2 = -\frac{\mu}{m_2}\dot{\boldsymbol{r}},
\end{aligned}
\end{equation}
where $\mu = \frac{m_1 m_2}{m_1 + m_2}$ denotes the reduced mass and $\boldsymbol{r} = \boldsymbol{x}_1 - \boldsymbol{x}_2$ is the relative position vector of the nuclei.
We propagate the inter-nuclear distance and the velocities of the two nuclei $\boldsymbol{v}_1$ and $\boldsymbol{v}_2$ employing the velocity-Verlet algorithm~\cite{Verlet1967,Mountney2025}. This algorithm recursively updates the inter-nuclear distance and the magnitude of the relative velocity at each time step as follows
\begin{equation}
\begin{aligned}
r_{n+1} &= r_n + v_n \Delta t + \frac{F_n}{2\mu} (\Delta t)^2, \\
v_{n+1} &= v_n + \frac{F_{n+1} + F_n}{2\mu} \Delta t.
\end{aligned}
\end{equation}
where $F_n$ is calculated from the PECs we obtained in Sec.~\ref{Sec::PECs_calculation} as follows
\begin{equation}
\begin{aligned}
F_n &= -\left.\frac{dU}{dr}\right|_{r = r_n},
\end{aligned}
\end{equation}
where $\Delta t$ is the time step in  the Monte-Carlo simulation. Following convergence tests, we adopt $\Delta t = 0.01~\text{fs}$.
 This computation allows us to obtain the velocities of the atomic fragments, according to Eq.~\eqref{eq:Velocities}, and hence to obtain the momenta and kinetic energy release (KER) at the asymptotic time limit, i.e. at the end of the time propagation.

\subsection{Initial conditions of the nuclei}

\begin{figure}[H]
    \centering
    \includegraphics[scale=1]{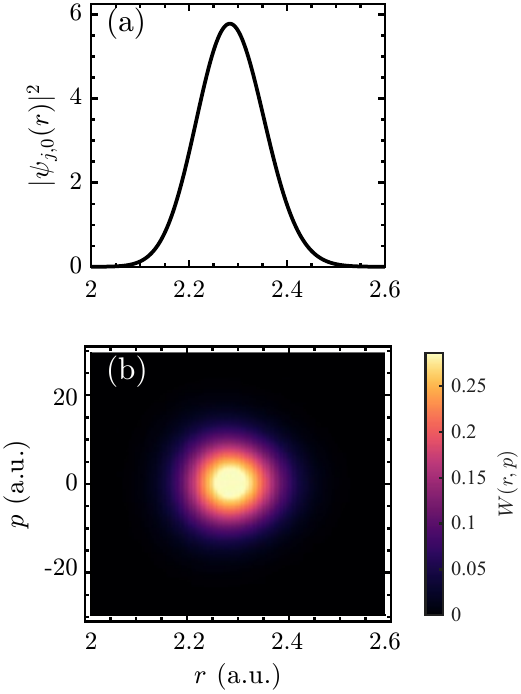}
    \caption{Top panel: distribution of the internuclear distance obtained from the squared Morse ground-state wave function of O$_2$. Bottom panel: Wigner function  of the O$_2$ ground state with respect to the inter-nuclear distance and relative momentum of the two nuclei. For each initial inter-nuclear distance $r_0$ sampled from the distribution in panel (a), the corresponding initial relative momentum $p_0$ is sampled from the conditional distribution defined by the slice of the Wigner function at  $r=r_0$ in panel (b).}
    \label{fig:Morse}
\end{figure}

In the Monte-Carlo simulation, for each event,  we need to determine the initial positions and momenta of the nuclei. We do so, using an importance sampling scheme, described in Ref.~\cite{Mountney2025}. Specifically, the initial inter-nuclear distance $r_0$ of the nuclei is sampled from the probability distribution given by the squared modulus of the Morse wave function of the ground state of neutral O$_2$, given in Ref. \cite{Frank2000MorseWigner} as follows 
\begin{equation}
\begin{aligned}
\psi_{j,0}(r) &= N_{j,0} e^{-\xi/2} \, \xi^{j} L_0^{2j}(\xi), \\
N_{j,0} &= \sqrt{\frac{2\beta j}{\Gamma(2j + 1)}}, \\
\xi &= (2j + 1)e^{-\beta |r - r_e|},\\
\omega_e &= \beta \sqrt{\frac{2D_e}{\mu}},
\end{aligned}
\end{equation}
where $\omega_e$ is the vibration frequency of the nuclei and $D_e$ is the dissociation energy of the ground state of neutral O$_2$. $L_0^{2j}(\xi)$ is an associated Laguerre polynomial~\cite{Abramowitz1965} and $\Gamma$ is the Gamma function.
 Next, for a given $r_0$, the corresponding Wigner function is evaluated to obtain the probability distribution of the relative  momentum of the two nuclei. Using this distribution we sample the initial momentum $p_0$ of the nuclei. The Wigner function for the ground state of neutral O$_2$ is given by
\begin{equation}
\label{eq:Wigner function}
W(\psi_{j,0} \mid r, p) = \frac{2}{\pi \Gamma(2j)} \, \xi^{2j} \, K_{-2ip/\beta}(\xi)
\end{equation}
where $K_{-2ip/\beta}(\xi)$ is the modified Bessel function of the third kind. In Fig.~\ref{fig:Morse}, we plot the distributions we use to sample the initial conditions in the Monte-Carlo simulations.

\label{Sec::Sampling}

\subsection{Monte Carlo simulation}
\label{Sec::Monte_Carlo_simulation}
In the Monte-Carlo simulation, once we define the initial inter-nuclear distance and the relative momentum of the nuclei, we propagate in time  the interaction of  O$_2$ with the XFEL-pulse. The time propagation is performed using a default time step of $\Delta t = 0.01$ fs. Each Monte-Carlo event  is initialized at 500 $\mathrm{fs}$ before the peak intensity of the pulse and propagated until 3000 $\mathrm{fs}$ after the peak intensity, ensuring sufficient time for all Auger-Meitner processes to occur. At each time step, we calculate the transition rates for the SPI and AM  processes  from a given  initial state to all energetically allowed  final states 
\begin{equation}
\begin{aligned}
\omega_{\alpha i}(t) &= \sigma_{\alpha i} J(t) \qquad \text{Photo-ionization} \\
\omega_{\alpha i}(t) &= \Gamma_{\alpha i} \qquad \text{Auger--Meitner decay},
\end{aligned}
\end{equation}
with $\sigma_{\alpha i}$ and $\Gamma_{\alpha i}$ being the SPI  cross sections and AM rates, respectively,  from initial state $\alpha$ to final  state i, and $J(t)$ being the photon flux.  For the molecular states, the SPI cross sections and AM rates  are evaluated at each instantaneous inter-nuclear distance  in the Monte-Carlo simulation.  We calculate these transition rates for the molecular ions as well as  for the atomic ion fragments, once  we transition from  the molecule to the two atoms.

At each time step, we stochastically determine which electronic transition takes place. The populations of the molecular and atomic states are assumed to evolve according to an exponential decay law~\cite{Jurek2016XMDYN}
\begin{equation}
P = P_0 e^{-\omega_{\alpha i} t}
\end{equation}
\begin{equation}
\label{eq:transition_time}
t_{\alpha i}(t) = -\,\frac{\ln \left(\frac{P}{P_0}\right)}{\omega_{\alpha i}}.
\end{equation}
We randomly generate the value of $\dfrac{P}{P_0}$ from 0 to 1, and  calculate the ``transition time" using Eq.~\eqref{eq:transition_time} for each possible transition from state $\alpha$ to state i. We identify the smallest time $t_{\alpha i}$, which corresponds to the most probable transition at the present time step. If this time is greater than the default time step of 0.01 fs, no electronic transition occurs and the time increases by the default time step of 0.01 fs. If $t_{\alpha i}$ is smaller than the default time step, then the time increases by $t_{\alpha i}$, and the transition to state i takes place. We update the corresponding molecular state and propagate the nuclei on the resulting PEC of the updated molecular state. 

Once we reach the distance where we transition from the molecule to the atomic fragments, we continue tracking the dynamics of the two nuclei and determine the transitions occurring in the resulting atomic fragments. At each time step, we track the transitions in the atomic fragments following  Eq.~\eqref{eq:transition_time}. At each time step, if no transition occurs in both atomic fragments, we take a $\Delta t = 0.01$ $\mathrm{fs}$ time step and update the inter-nuclear distance and the relative velocity of the nuclei. If a transition occurs in one of the atomic fragments, we increase the charge of that fragment by one and calculate the new Coulomb force with the updated charges in the time interval $t_{\alpha i}$. If transitions occur in  both atomic fragments at a given time step, we first increase the charge of the atomic fragment where the earliest electronic transition takes place and update the inter-nuclear distance, velocities and Coulomb force up until t+$t_{\alpha i}^{1}$. Then, from time  t+$t_{\alpha i}^{1}$ to time t+$t_{\alpha i}^{1}$+$t_{\alpha i}^{2}$, with $t_{\alpha i}^{2}>t_{\alpha i}^{1}$,  we increase the charge of the other atomic ion as well and compute the new inter-nuclear distance and velocities. We continue this process until 3000 fs after the peak intensity of the laser pulse. For each  Monte-Carlo event, we record all molecular and atomic ion states, the time and inter-nuclear distance when each electronic  transition occurs, 
the probability for a certain fragmentation channel to occur as well as the sum of the kinetic energies of the atomic fragments  (KER) for each channel. We also record the sequences of SPI and AM transitions, i.e. the ionization pathways that  lead to each fragmentation channel.

\section{Results}\label{Sec::Result}

\begin{figure}[H]
    \centering
    \includegraphics[scale=0.8]{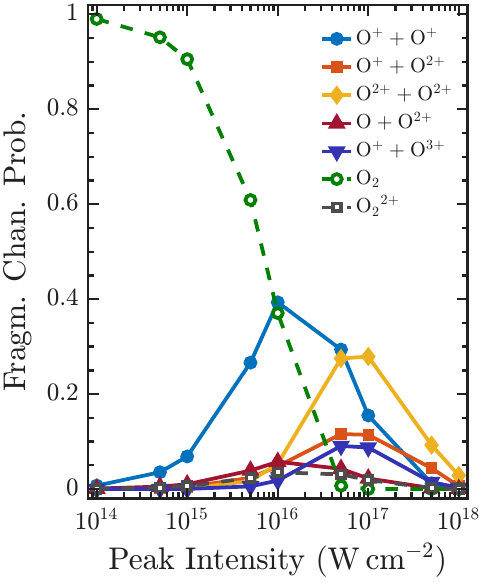}
    \caption{ Out of all Monte-Carlo events, probability  as a function of the pulse  intensity for the fragmentation channels  with total charge of the  fragments  up to 4 (solid lines) and for non-dissociating molecular ion states (dashed lines) to result following the interaction of O$_{2}$ with the XFEL-pulse. The laser intensity is shown  on a logarithmic scale. Symbols denote the intensities where we perform  Monte-Carlo simulations,  while the  connecting lines are included for guiding the eye.}
            \label{fig:intensity}
\end{figure}

In what follows, we compute the KER for different fragmentation channels, identify the pathways contributing to these channels, and explain the origin of the peaks in the KER spectra. In Fig.~\ref{fig:intensity}, we plot the probability, out of all Monte-Carlo events, for fragmentation channels to result following the interaction of O$_2$ with the XFEL-pulse as well as the probability  for molecular ion states to withstand the interaction with the XFEL-pulse.  Since we account for up to O$_{2}^{5+}$ molecular ion states, we consider fragmentation channels where the sum of the charges of the atomic ion fragments is up to 4. Moreover, to accommodate the general expectation for a rapid charge up of a molecule in CEI, we consider a pulse duration of 10 fs. However, we find that our findings for the KER are robust and hold for shorter and longer pulses. We choose a 10 fs pulse, since it is a short enough pulse, while there is no need to account for the bandwidth of the pulse.  

 We choose the intensity of the pulse to be 5$\times 10^{16}$ Wcm$^{-2}$, since Fig.~\ref{fig:intensity} shows that at this intensity the probability for all resulting fragmentation channels is significant.  We find that the O$^{+}$+O$^{+}$ and   O$^{2+}$+O$^{2+}$ channels are the most probable ones. In what follows, we discuss how extraction of the equilibrium (initial) inter-nuclear distance of the ground state of O$_{2}$ is more readily obtained from the KER spectra of the  O$^{+}$+O$^{+}$  channel rather than the O$^{2+}$+O$^{2+}$ channel. This is  contrary to general expectation. Indeed, the former channel reaches Coulombic behaviour at larger distances than the ones  for the latter channel. Hence, according to  CEI, one would expect that it is  the KER of the latter channel and not the former one that allows for an accurate determination of 2.296 a.u., the initial inter-nuclear distance of O$_{2}$. 

\subsection{KER of the O$^{+}$ + O$^{+}$ Channel}

We  consider the fragmentation channel O$^{+}$+O$^{+}$ resulting from the interaction of O$_{2}$ with the XFEL-pulse. We find that  P$_{C}$A$_{CV}$ is the dominant pathway leading to this channel. That is, first, a single photon is absorbed from a core molecular orbital (C), either from the $\mathrm{1\sigma_{g}}$ or the $\mathrm{1\sigma_{u}}$ orbital, and, then, an AM process (A$_{CV}$) takes place, with a valence electron filling the core hole and another valence electron taking the excess energy released  and escaping to the continuum. We find (not shown) that these two transitions occur at inter-nuclear distances close to the initial inter-nuclear distance of O$_{2}$, which is 2.296 a.u.. 

\begin{figure}[H]
    \centering
    \includegraphics[scale=0.8]{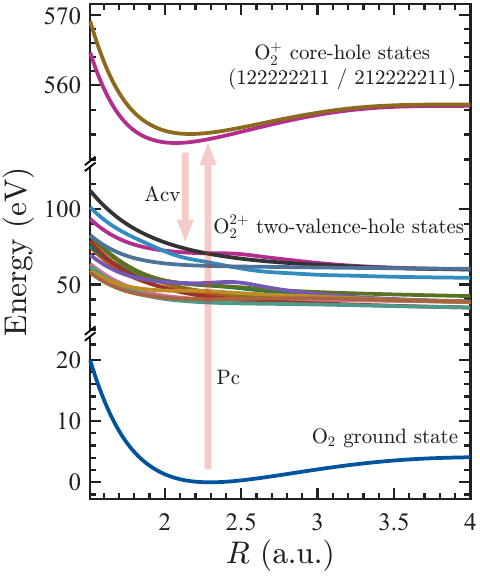}
    \caption{PECs involved in the schematic representation of the P$_{C}$A$_{CV}$ pathway leading to the O$^{+}$ + O$^{+}$ fragmentation channel, see text for details. Both the P$_{C}$ and A$_{CV}$ processes occur around the initial distance of the ground state of O$_{2}$.}
        \label{fig:PECsforO+O+}
\end{figure}

In this channel, we find that in  the P$_{C}$ process, a transition  occurs from the ground state 222222211 to the 122222211 or  212222211 O$_{2}^{+}$ core-hole molecular states,
with the hole in the $\mathrm{1\sigma_{g}}$ or  $\mathrm{1\sigma_{u}}$ orbital, respectively. In the AM process, the transition takes place from the two intermediate core-hole states to 14 O$_{2}^{2+}$ molecular states with two valence electrons missing, see Fig. \ref{fig:PECsforO+O+}. That is, 
following A$_{CV}$, the molecule transitions to one of the 14  two-valence-hole molecular states, which then dissociate leading to two O$^{+}$ fragments.

We find that these 14 final O$_{2}^{2+}$  two-valence-hole molecular states  play a crucial role in the KER spectra. Indeed, Fig.~\ref{fig:figKERO+O+} A shows the KER spectra  of  the O$^{+}$ + O$^{+}$ channel (top row). Fig.~\ref{fig:figKERO+O+} B and sub-plots (a)-(r) (on the left)   show the contribution to the KER spectra of each of the 14 O$_{2}^{2+}$ states.  The  KER of each final state are presented   in descending order of the \% contribution of each state to the probability of the O$^{+}$ + O$^{+}$  channel. Also, per row in Fig.~\ref{fig:figKERO+O+} B, on the right, we plot the corresponding PECs for the three O$_{2}^{2+}$ states whose contribution to the KER we plot on the left.

Comparing the peaks of the KER in Fig.~\ref{fig:figKERO+O+} A with the peaks of the KER of each final O$_{2}^{2+}$  two-valence-hole molecular state in Fig.~\ref{fig:figKERO+O+} B, we find a  clear correspondence.  Indeed, the 7.6 eV peak  in the KER in A is due to the 7.6 peak in the KER of the 221222201 / 221222210 states in sub-plot (q), the 8.6 eV peak  in the KER in A is due to the 8.6 eV peak in the KER of the 222221201 / 222222110 states in sub-plot (j), the 9.9 eV peak in the KER in A is mainly due to the 9.9 eV peak of the KER of the  222221210 / 222222101 states in sub-plot (m) and  the 9.4 eV peak in the KER  of the states in  sub-plot (r). Also, the shoulder around 12.4 eV in the KER in A is mainly due to the 12.1 eV peak in the KER of the states in sub-plot (a) as well as of the 13.4 eV peak in the KER of the states in sub-plot (e). The 14.4 eV peak in the KER in A is due to the 14.4 eV peak in the KER of the states in sub-plots  (c) and  (f) as well as of the 14.1 eV peak in the KER of the states in sub-plots (k) and (o). The broad 17.9 eV peak in the KER in A is mainly due to the 17.9 peak in the KER of the states in sub-plot (b) and the 17.4 eV peak of the state  in sub-plot (g). Finally, the 20.1 eV peak  in the KER in A is due to the 20.1 eV peak in the KER of the states in sub-plot (n).

\begin{widetext}
\begin{center}
    \includegraphics[scale=0.93]{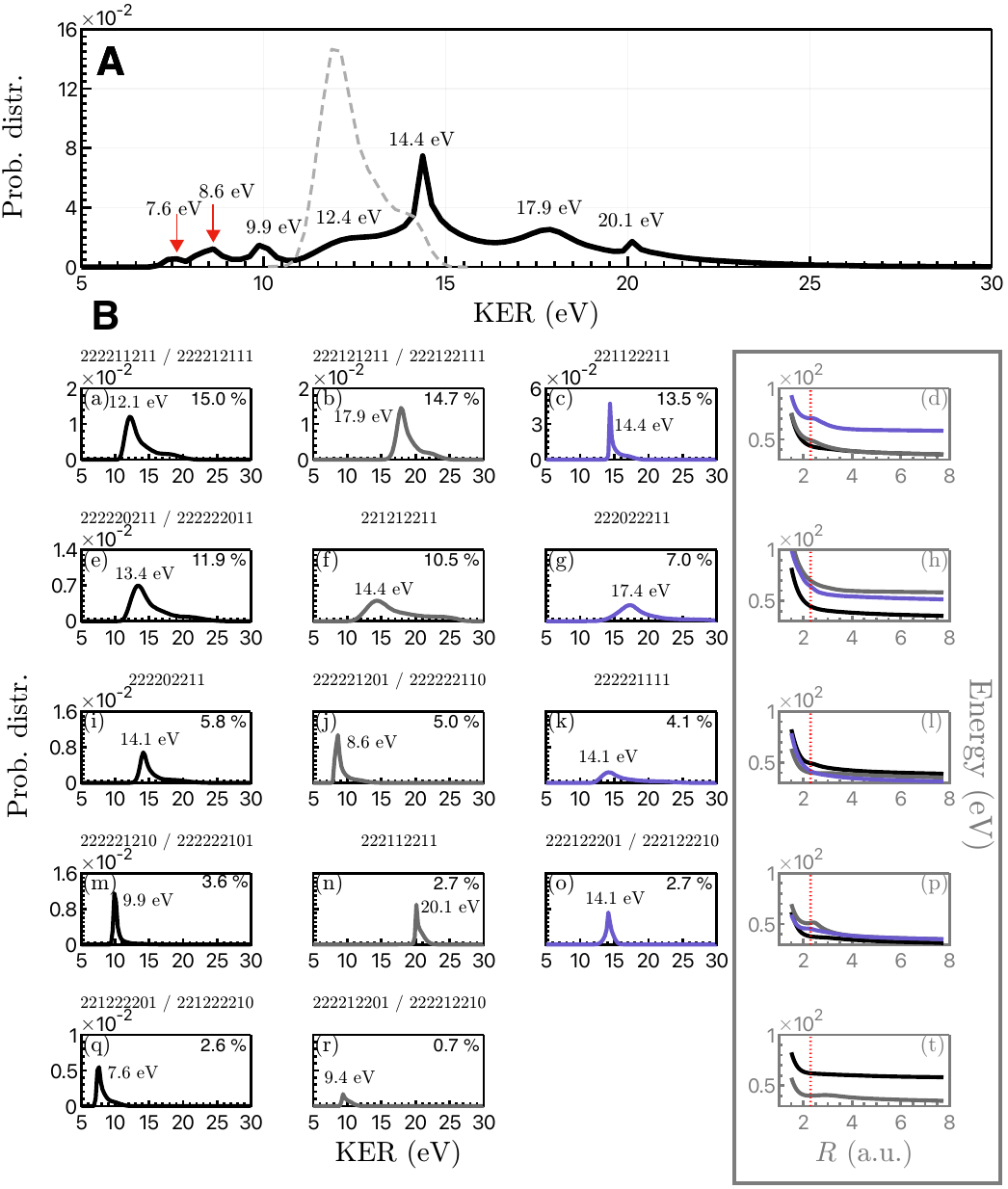}

    \captionof{figure}{Top row A: we plot the KER for the O$^{+}$ + O$^{+}$ fragmentation channel (black line). The gray line shows the KER one would obtain if Coulomb explosion was to occur at the inter-nuclear distances (not shown) at the time the A$_{CV}$ process occurs. Also, we show the peaks in the KER spectra. In B, we show on the left the contribution of each of the final 14 O$_2^{2+}$ two-valence-hole molecular states to the KER spectra. We also identify the peak of the KER of each final O$_2^{2+}$ state as well as the \% contribution of this state to the formation of the O$^{+}$ + O$^{+}$ channel. For each row in B, on the right, we plot the PECs for the three O$_2^{2+}$ states whose KER we plot on the left. To distinguish the PEC on the right that corresponds to the KER of a final state on the left we use three different colors, black, gray and blue. The dotted red line denotes the initial inter-distance of O$_2$. The / symbol in the states at the top of each sub-plot denotes degenerate molecular states.}
    \label{fig:figKERO+O+}
\end{center}
\end{widetext}

\begin{figure}[H]
    \centering
    \includegraphics[scale=1.0]{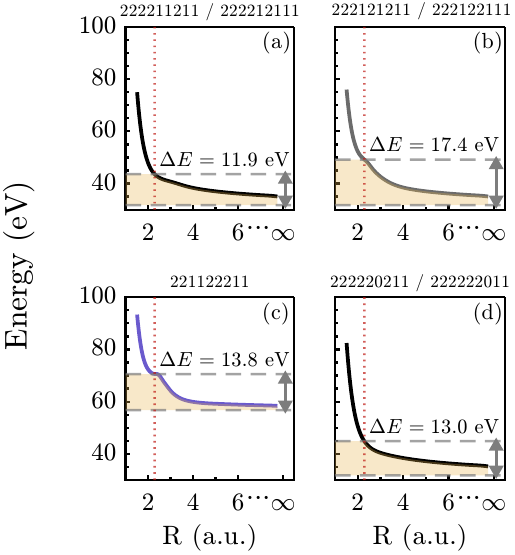}
    \caption{We show how the peak value in the KER of  the  final O$_{2}^{2+}$ states in subplots (a), (b), (c) and (e) in Fig.~\ref{fig:figKERO+O+} B is obtained by subtracting from the energy of the PEC at the  2.296 a.u. initial inter-nuclear distance of O$_2$ (red dotted line) the asymptotic value of the PEC, i.e. the dissociation energy limit of the molecular state.}
    \label{fig:KERdiscuss}
\end{figure}

We find that the peak in the KER corresponding to each final O$_{2}^{2+}$ state, depicted in the sub-plots in  Fig.~\ref{fig:figKERO+O+} B, is obtained by subtracting from the energy of the respective PEC (plotted on the right in Fig.~\ref{fig:figKERO+O+} B)  at the 2.296 a.u. equilibrium (initial) inter-nuclear distance of O$_2$ the asymptotic energy of this PEC, i.e. the dissociation energy limit of the molecular state.  That is, we obtain the peak of the KER of each final O$_{2}^{2+}$ state by conservation of potential and kinetic energy, assuming zero kinetic energy at the equilibrium inter-nuclear distance. This is depicted  in   Fig.~\ref{fig:KERdiscuss}.

 We note that  the KER of the O$^{+}$ + O$^{+}$ fragmentation channel are not consistent with Coulomb explosion spectra. Indeed, in Fig.~\ref{fig:figKERO+O+} A, along with the KER for the O$^{+}$+O$^{+}$ channel, we plot the KER one would obtain by Coulomb repulsion using the inter-nuclear distances where the A$_{CV}$ process takes place (not shown). We obtain this distribution of inter-nuclear distances from our Monte-Carlo simulation. A comparison of the KER obtained with the Monte-Carlo simulation and the one obtained from just the Coulomb repulsion at the inter-nuclear distances where the A$_{CV}$ process occurs clearly shows that they are quite different. This is expected, since for roughly  30\% of the probability of the O$^{+}$ + O$^{+}$ channel, the final O$_{2}^{2+}$ molecular PECs are mostly influenced by chemical bonding and only at large distances they follow Coulomb repulsion curves. In addition, the inter-nuclear distances where the P$_{C}$ and the A$_{CV}$ processes occur 
vary from  1.8 a.u. to 2.5 a.u., i.e small inter-nuclear distances where the PECs are mostly influenced by chemical bonding.

So far we have found a clear correspondence  of the peaks of the KER of the  O$^{+}$ + O$^{+}$ fragmentation channel   in Fig.~\ref{fig:figKERO+O+} A with the peaks of the KER of the O$_{2}^{2+}$ two-valence-hole states, the molecule proceeds to dissociate from,  in Fig.~\ref{fig:figKERO+O+} B. The above, in conjunction with the fact that these peaks correspond to the difference in the energy of the PEC at the equilibrium inter-nuclear distance of O$_{2}$ and at the asymptotic limit, demonstrate how the KER spectra of this channel   image the equilibrium inter-nuclear distance of the molecule. They do so without employing CEI, since the KER spectra are not consistent with Coulomb repulsion.

We further make  the point of how the KER spectra image the  2.296 a.u. equilibrium inter-nuclear distance in Table~\ref{tab:exact_r_11}. We follow   the inverse procedure than the one depicted in  Fig.~\ref{fig:KERdiscuss}, that is, we identify the inter-nuclear distance 
that corresponds to the energy of the final state that is obtained by adding to the peak energy of each final O$_{2}^{2+}$ state in Fig.~\ref{fig:figKERO+O+} B the dissociation limit of the corresponding  PEC.  As a result, we obtain the inter-nuclear distances in Table~\ref{tab:exact_r_11}, with several of them being very close to the  2.296 a.u. equilibrium  inter-nuclear distance of O$_{2}^{2+}$. 

Experimental identification  of the KER of  each of these  14 O$_{2}^{2+}$ states could be  potentially realised by measuring the A$_{CV}$ electron and the atomic ions fragments in coincidence. As Table~\ref{tab:exact_r_22} shows, the A$_{CV}$ electron energies in the O$^+$+O$^{+}$ channel are quite distinct for each final O$_{2}^{2+}$ two-valence hole molecular state. The usefulness of measuring an escaping electron and ions in coincidence for interaction of molecules with X-ray pulses has been previously addressed in Ref.~\cite{ref19,ref4}.

\begin{table}[H]
\centering
\footnotesize
\caption{We show the peak positions of the KER of the O$_{2}^{2+}$ two-valence-hole molecular states in Fig.~\ref{fig:figKERO+O+} B and, following the inverse procedure than the one depicted in  Fig.~\ref{fig:KERdiscuss}, we identify the corresponding initial inter-nuclear distance.}
\label{tab:exact_r_11}
\begin{tabular}{lcc}
\multicolumn{3}{c}{
}
  \\
\hline
Final O$_{2}^{2+}$ state & Peak Energy (eV) & $r$ (a.u.) \\
\hline
\texttt{222211211 / 222212111} & 12.1 & 2.27 \\
\texttt{222121211 / 222122111} & 17.9 & 2.24 \\
\texttt{221122211} & 14.4 & 2.13 \\
\texttt{222220211 / 222222011} & 13.4 & 2.27 \\
\texttt{221212211} & 14.4 & 2.24 \\
\texttt{222022211} & 17.4 & 2.23 \\
\texttt{222202211} & 14.1 & 2.25 \\
\texttt{222221201 / 222222110} & 8.6 & 2.22 \\
\texttt{222221111} & 14.1 & 2.27 \\
\texttt{222221210 / 222222101} & 9.9 & 2.30 \\
\texttt{222112211} & 20.1 & 1.97 \\
\texttt{222122201 / 222122210} & 14.1 & 2.03 \\
\texttt{221222201 / 221222210} & 7.6 & 1.98 \\
\texttt{222212201 / 222212210} & 9.4 & 2.00 \\
\hline
\hline
\end{tabular}
\end{table}

\begin{table}[H]
\centering
\footnotesize
\caption{
A$_{CV}$ electron energies for the O$^{+}$ + O$^{+}$ fragmentation channel for each of the 14 final O$_{2}^{2+}$ states in Fig.~\ref{fig:figKERO+O+} B  when the electron first transitions via a P$_{C}$ process to one of the two  core-hole-states, i.e. the 122222211 or the 212222211.
}
\label{tab:exact_r_22}
\begin{tabular}{lc}
\multicolumn{2}{c}{
} \\
\hline
Final O$_{2}^{2+}$ state & A$_{CV}$ electron energy  (eV)   \\
\hline
\texttt{222211211 / 222212111} & 509 \\
\texttt{222121211 / 222122111} & 503 \\
\texttt{221122211} & 482 \\
\texttt{222220211 / 222222011} & 507 \\
\texttt{221212211} & 482 \\
\texttt{222022211} & 488\\
\texttt{222202211} & 503 \\
\texttt{222221201 / 222222110} & 512 \\
\texttt{222221111} & 511 \\
\texttt{222221210 / 222222101} & 514 \\
\texttt{222112211} & 501 \\
\texttt{222122201 / 222122210} & 507 \\
\texttt{221222201 / 221222210} & 490 \\
\texttt{222212201 / 222212210} & 512 \\
\hline
\hline
\end{tabular}
\end{table}

\subsection{KER of the O$^{+}$ + O$^{2+}$ Channel}

Next, we  consider the O$^{+}$+O$^{2+}$ fragmentation channel  resulting from the interaction of O$_{2}$ with the XFEL-pulse. We find that  P$_{C}$A$_{CV}$P$_{V}$, P$_{C}$P$_{V}$A$_{CV}$, P$_{V}$P$_{C}$A$_{CV}$  are the dominant pathways leading to this channel, with P$_{V}$ denoting single photon absorption from a valence molecular orbital. That is, three sequential processes take place. One involves single photon absorption from a core orbital, another one an  AM process where a valence electron fills in a core hole and another valence electron escapes as well as  another process that involves a single photon absorption process from a valence orbital.  
 We find (not shown) that the first  two transitions occur at inter-nuclear distances close to the equilibrium  inter-nuclear distance of O$_{2}$, which is 2.296 a.u. However,  the last transition occurs at distances that vary from being  close to the equilibrium inter-nuclear distance to much larger distances. 

\begin{figure}[H]
    \centering
    \includegraphics[scale=1.0]{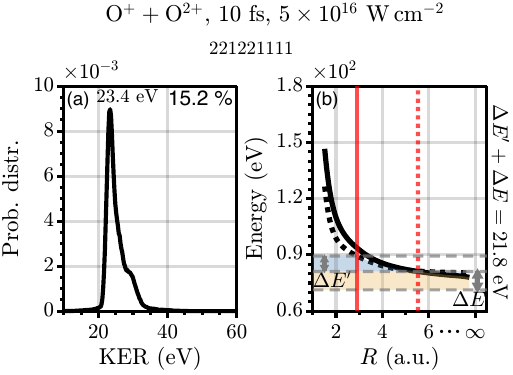}
    \caption{We plot in (a) the KER spectra of the 221221111 final O$_{2}^{3+}$ state, which differs from the other final O$_{2}^{3+}$ states in Fig.~\ref{fig:KERO++O2+} B  by the pathway that leads to the O$^{+}$+O$^{2+}$ channel and has two AM transitions instead of one for all other final O$_{2}^{3+}$ states. }
    \label{fig:KERonly}
\end{figure}

In this fragmentation channel, we find that, following the first two processes in the dominant pathways, the molecule transitions to several intermediate states with a core and valence hole or just two valence holes. This differs from the O$^{+}$+O$^{+}$ channel, where only two core-hole intermediate states are reached.   The intermediate  states  in the O$^{+}$+O$^{2+}$ channel are different for each of the final O$_{2}^{3+}$ molecular states that are reached following the third transition. Also, in the O$^{+}$+O$^{2+}$ channel, there are significantly more final states than the 14 ones in the O$^+$+O$^+$ channel.  Indeed, the 14 final states of  the O$^+$+O$^{2+}$ channel shown in  Fig.~\ref{fig:KERO++O2+} B and the final state  shown in Fig.~\ref{fig:KERonly} account for roughly 74\% of the probability for the O$^{+}$+O$^{2+}$ channel to occur. There are  several more final O$_{2}^{3+}$ states contributing less than 1\% each to the formation of O$^+$+O$^{2+}$. The final state of O$^{3+}_{2}$ in Fig.~\ref{fig:KERonly}  differs from the ones in Fig.~\ref{fig:KERO++O2+} B in terms of the three-process pathway  P$_{C}$A$_{CV}$A$_{VV}$, with one single photon and two AM transitions occurring in this case. 

Comparing the peaks of the KER in Fig.~\ref{fig:KERO++O2+} A with the peaks of the KER of each final O$_{2}^{3+}$ molecular state in Fig.~\ref{fig:KERO++O2+} B, we  find that there is not as  clear a correspondence as for the O$^{+}$+O$^{+}$ channel.  Indeed, 
the KER spectra in Fig.~\ref{fig:KERO++O2+} A for all events leading to the O$^{+}$+O$^{2+}$ channel have significantly less well defined peaks compared to the KER of  the O$^{+}$+O$^{+}$ channel. Also, the KER spectra of each individual final O$_{2}^{3+}$ state in Fig.~\ref{fig:KERO++O2+} B 
are significantly wider compared to the KER spectra of the  O$_{2}^{2+}$ states in Fig.~\ref{fig:figKERO+O+} B.  Nevertheless, there is still some correspondence between the total KER spectra and the KER spectra of the individual final states. Such is the case  for the 25.6 eV peak in  Fig.~\ref{fig:KERO++O2+} A and sub-plot (n) in Fig.~\ref{fig:KERO++O2+} B, for the 27.4 eV peak and sub-plot (b), for the 33.1 eV peak and sub-plot (a), as well as the 40.1 eV peak and sub-plot (c). Moreover, the 23.4 eV peak in the total KER spectra in Fig.~\ref{fig:KERO++O2+} A corresponds to the 23.4 eV peak in Fig.~\ref{fig:KERonly} (a).

\begin{widetext}
\begin{center}
   
  \includegraphics[scale=0.75]{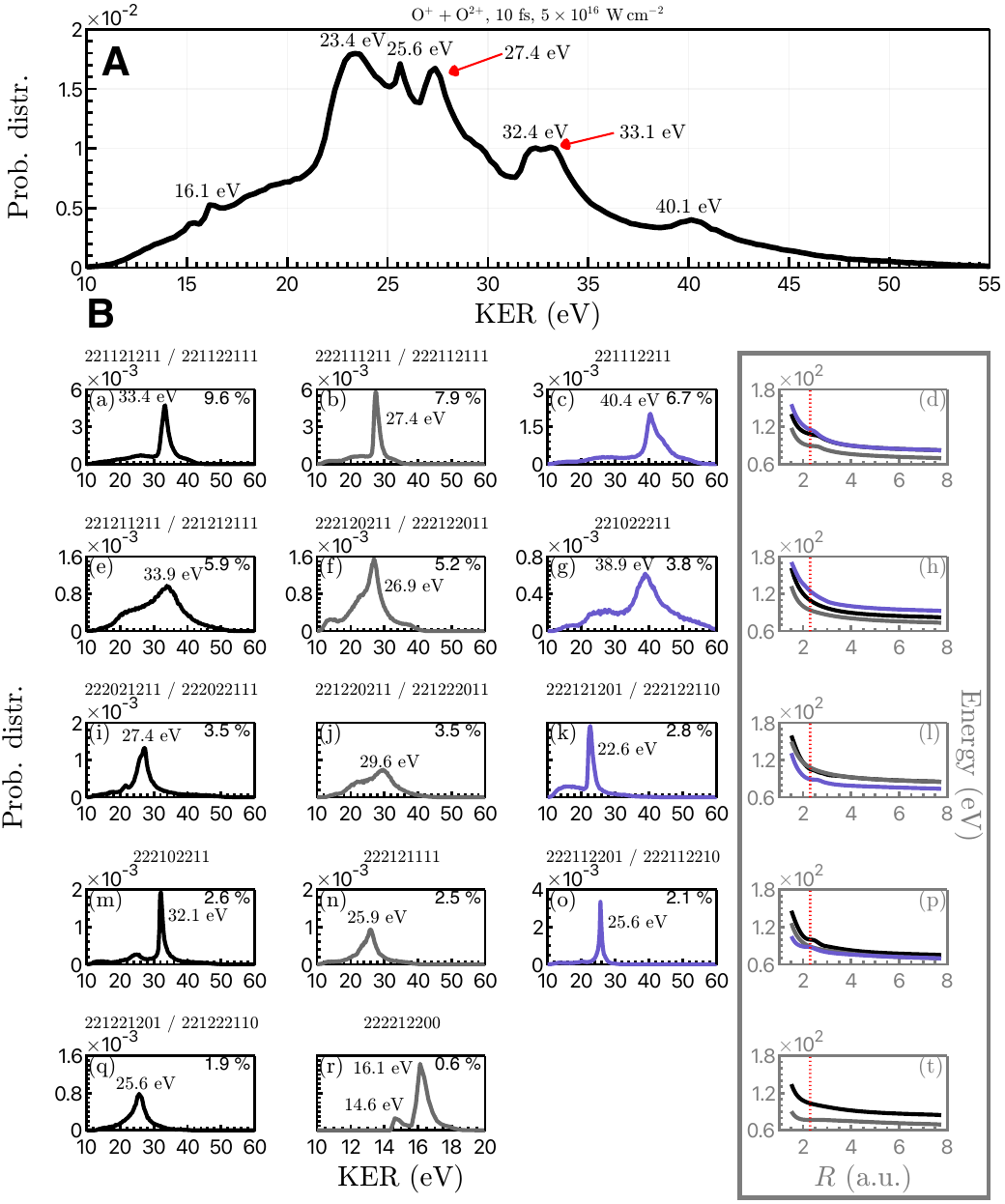}
    \captionof{figure}{As in Fig.~\ref{fig:figKERO+O+} A and Fig.~\ref{fig:figKERO+O+} B for the O$^+$+O$^{2+}$ channel. }
    \label{fig:KERO++O2+}
\end{center}
\end{widetext}

The KER spectra of the individual O$_{2}^{3+}$ states in Fig.~\ref{fig:KERO++O2+} B are not only wider but have in many cases a broad secondary peak at smaller values of the KER. The sharper peaks at larger values of the KER of the final states in Fig.~\ref{fig:KERO++O2+} B are obtained 
in the same way as we obtain the peaks for the O$^+$+O$^{+}$ channel. Namely,  we subtract the energy of the PEC of the final O$_{2}^{3+}$ state from the dissociation energy limit, obtained by the PEC   at an asymptotic distance. Following the inverse procedure, we identify the inter-nuclear distance 
that corresponds to the energy of the final O$_{2}^{3+}$ state that is obtained by adding to the peak energy of each final O$_{2}^{3+}$ state in Fig.~\ref{fig:KERO++O2+} B the dissociation limit of the corresponding  PEC.  As a result, we extract the inter-nuclear distances in Table~\ref{tab:exact_r_12}, with several of them being very close to the  2.296 a.u. equilibrium  inter-nuclear distance of O$_2$.  This is similar to the inverse process applied for the O$^{+}$+O$^{+}$ channel and obtaining the distances in Table~\ref{tab:exact_r_11}.

\begin{table}[H]
\centering
\footnotesize
\caption{We show the sharper peak energies of the KER of the dissociating O$_{2}^{3+}$  molecular states in Fig.~\ref{fig:KERO++O2+} B and extract the corresponding inter-nuclear distance of O$_2$, see text.}
\label{tab:exact_r_12}
\begin{tabular}{lcc}
\multicolumn{3}{c}{
} \\
\hline
Final O$^{3+}_{2}$ state & Peak (eV) & $r$ (a.u.) \\
\hline
\texttt{221121211 / 221122111} & 33.4 & 2.23 \\
\texttt{222111211 / 222112111} & 27.4 & 2.22 \\
\texttt{221112211}              & 40.4 & 2.26 \\
\texttt{221211211 / 221212111} & 33.9 & 2.31 \\
\texttt{222120211 / 222122011} & 26.9 & 2.30 \\
\texttt{221022211}              & 38.9 & 2.29 \\
\texttt{222021211 / 222022111} & 27.4 & 2.29 \\
\texttt{221220211 / 221222011} & 29.6 & 2.31 \\
\texttt{222121201 / 222122110} & 22.6 & 2.25 \\
\texttt{222102211}              & 32.1 & 2.21 \\
\texttt{222121111}              & 25.9 & 2.30 \\
\texttt{222112201 / 222112210} & 25.6 & 2.04 \\
\texttt{221221201 / 221222110} & 25.6 & 2.30 \\
\texttt{222212200}              & 16.1 & 1.84 \\
\hline
\hline
\end{tabular}
\end{table}


However, the peak of the KER spectra of the 221221111 O$_{2}^{3+}$ state in
Fig.~\ref{fig:KERonly} (a) is not obtained by subtracting from the energy of the corresponding PEC at  $2.296~\mathrm{a.u.}$ the dissociation energy limit. This would yield 30.7 eV which is much higher than the actual 23.4 eV peak. In Fig.~\ref{fig:KERonly} (b), we show that to obtain the 23.4 eV peak 
we first identify where the inter-nuclear distribution of the last electronic transition peaks in the A$_{VV}$ process of the  P$_{C}$A$_{CV}$A$_{VV}$ pathway. We find this distance to be 5.54 a.u. Then,  we compute $\Delta E$ by subtracting from the energy of the respective PEC at 5.54 a.u. the dissociation energy limit. Moreover, at 5.54 a.u.,  the dissociating molecule has already acquired kinetic energy while moving on the intermediate molecular states, denoted by the black dotted PEC in Fig.~\ref{fig:KERonly} (b). These intermediate molecular states are reached after two electronic transitions which take place close to the equilibrium inter-nuclear distance. Hence, to obtain the kinetic energy acquired $\Delta E'$, we subtract from the energy of the PEC of an intermediate state  at 2.296 a.u. the energy  at 5.54 a.u. However, there are several intermediate molecular states. Hence we need to average the differences  $\Delta E'$ using as weights the probability of each intermediate state to be reached by electronic transitions. As a result, we obtain $\Delta E' + \Delta E = 21.8~\mathrm{eV}$, which is relatively close to the actual peak at 23.4 eV (Fig.~\ref{fig:KERonly} (a)).

As we have already mentioned, for the KER of the final O$_{2}^{3+}$ states  in Fig.~\ref{fig:KERO++O2+} B there are two peaks, a sharp one at higher energies and a broader one at smaller energies. Next, we demonstrate how to obtain these  broader peaks following a procedure similar to the one 
followed to obtain the peak in the KER of the  221221111 O$_{2}^{3+}$ state in Fig.~\ref{fig:KERonly} (a). For instance, we consider the broad peak in the KER of the sub-plot (a) in  Fig.~\ref{fig:KERO++O2+} B.  We find that the distribution of the inter-nuclear distances, where the final electronic transition occurs, peaks at 2.98 a.u. for KER in the energy range from 10 eV to 30 eV (not shown).  We then find that $\Delta E$, computed by subtracting from the energy of the 221221111  PEC at 2.98 a.u. the asymptotic energy of the PEC, i.e. the dissociation energy limit, is equal to 23.35 eV. Also, the kinetic energy acquired 
at the inter-nuclear distance 2.98 a.u. is due to the molecule dissociating via the intermediate molecular states that are reached after two electronic transitions. These transitions occur at inter-nuclear distances close to 2.296 a.u. Hence, to compute $\Delta E'$, we subtract from the energy of the intermediate-state PEC  at 2.296 a.u. the energy of the PEC  at 2.98 a.u. Since there are several intermediate molecular states, we  average the different  $\Delta E'$ using as weights the probability for each intermediate state to be reached after two electronic transitions.
 These intermediate states are shown in Table~\ref{tab:exact_r_12}.  
As a result, we obtain $\overline{\Delta E'} + \Delta E = 23.35+3.68=27.03 ~\mathrm{eV}$, which is close to the broad peak value of the KER shown in sub-plot (a) in Fig.~\ref{fig:KERO++O2+} B. Following a similar procedure, we find that  $\overline{\Delta E'} + \Delta E = 21.58+2.50=24.08  ~\mathrm{eV}$ for the sub-plot (b) in Fig.~\ref{fig:KERO++O2+} B,  where $\overline{\Delta E'}$ is obtained in Table ~\ref{tab:exact_r_12}.


\begin{table}[H]
\centering
\footnotesize
\caption{
$\Delta E'$ corresponding to the \texttt{221121211}/\texttt{221122111} states whose KER are shown in sub-plot (a) in Fig.~\ref{fig:KERO++O2+} B.}
\label{Table1}
\begin{tabular}{ccc}
\multicolumn{3}{c}{} \\
\hline
Intermediate O$_{2}^{2+}$ states & Probability & $\Delta E'$ (eV) \\
\hline
\texttt{221222111 / 221221211} & 0.2420 & 4.24 \\
\texttt{222122111 / 222121211} & 0.2065 & 6.33 \\
\texttt{122122211} & 0.1775 & 3.07 \\
\texttt{212122211} & 0.09332 & 3.26 \\
\texttt{121222211} & 0.09109 & 1.02 \\
\texttt{221122211} & 0.05786 & 6.16 \\
\texttt{122222111 / 122221211} & 0.05083 & -0.14 \\
\texttt{211222211} & 0.04845 & 0.84 \\
\texttt{212222111 / 212221211} & 0.03235 & 0.41 \\
\hline
 & & $\overline{\Delta E'} = 3.68$ \\
\hline
\hline
\end{tabular}
\end{table}

\begin{table}[H]
\centering
\footnotesize
\caption{$\Delta E'$ corresponding to the  \texttt{222111211}/\texttt{222112111} states whose KER are shown in sub-plot (b) in Fig.~\ref{fig:KERO++O2+} B.}
\label{Table2}
\begin{tabular}{ccc}
\multicolumn{3}{c}{} \\
\hline
Intermediate O$_{2}^{2+}$ states & Probability & $\Delta E'$ (eV) \\
\hline
\texttt{222212111 / 222211211} & 0.3192 & 2.49 \\
\texttt{122122211} & 0.2360 & 2.78 \\
\texttt{212122211} & 0.1384 & 2.96 \\
\texttt{122212211} & 0.07241 & -0.70 \\
\texttt{222122111 / 222121211} & 0.07048 & 5.50 \\
\texttt{222112211} & 0.04980 & 5.94 \\
\texttt{122222111 / 122221211} & 0.04811 & 0.01 \\
\texttt{212212211} & 0.03443 & -0.46 \\
\texttt{212222111 / 212221211} & 0.03114 & 0.53 \\
\hline
 & & $\overline{\Delta E'} = 2.50$ \\
\hline
\hline
\end{tabular}
\end{table}

\clearpage

Hence, comparing the KER of the O$^{+}$+O$^{+}$ channel in  Fig.~\ref{fig:figKERO+O+} A and B with the KER of the O$^{+}$+O$^{2+}$ channel in Figs~\ref{fig:KERO++O2+} A and B, the KER of the channel with the higher charge of the atomic fragments are wider and have less of a correspondence 
 with the peaks in energy of the KER of the individual final states. The reason is two-fold. On one hand, in the O$^{+}$+O$^{2+}$,  more intermediate O$_{2}^{2+}$ molecular states are reached after two electronic transitions  channel compared to just two intermediate O$_{2}^{+}$ states reached after 
one electronic transition 
in the  O$^{+}$+O$^{+}$  channel. Moreover, significantly more final O$_{2}^{3+}$ states are reached in the O$^{+}$+O$^{2+}$ channel, which the molecule follows to dissociate compared to the 14 final O$_{2}^{2+}$ states that are reached in the  O$^{+}$+O$^{+}$ channel, which  the molecule follows 
to dissociate. The larger number of intermediate and final states leads to overlapping peaks in the total KER of the  O$^{+}$+O$^{2+}$ channel.  Another significant difference between the two channels, is that in the lower charge channel the electronic transitions take place close to the equilibrium
 inter-nuclear distance of O$_{2}$. However, in the higher charge  channel the third transition takes place at significantly larger distances, resulting in a broad shoulder at the lower energies of the KER in addition to the sharp peak at the higher values of the KER, the latter peak associated with the equilibrium inter-nuclear distance, as we have already shown. These broad shoulders are due to the kinetic energy the atomic fragments acquire while dissociating on the molecular states reached one electronic transition less before the final molecular states are reached. These broad shoulders of the individual KER result in even broader total KER spectra. Given the above, it is clear that extracting the equilibrium distance from the KER of the lower charge channel is easier than from the higher one.

\subsection{O$^{2+}$ + O$^{2+}$ Channel}

In what follows, we consider the fragmentation channel O$^{2+}$+O$^{2+}$, resulting from the interaction of the O$_{2}$ molecule with the XFEL-pulse. We find that  P$_C$P$_C$A$_{CV}$A$_{CV}$  is the dominant pathway leading to the formation of the  O$^{2+}$+O$^{2+}$ channel.
 Namely, first, there are two sequential single photon absorptions from core electrons, followed by two sequential A$_{CV}$ processes. In each AM process, a valence electron fills in one core hole and another valence electron is ejected  to the continuum. We find that the first three electronic transitions occur at distances close to the equilibrium inter-nuclear distance of O$_{2}$. The last electronic process occurs at much larger distances, which on average are even larger than the average distance where the third electronic transition occurs in the O$^{+}$+O$^{2+}$ channel.  

In this fragmentation channel, we find that, following the first three processes in the dominant pathway, the molecule transitions to several intermediate O$_{2}^{3+}$ states with a core and two valence holes. This differs from the O$^{+}$+O$^{+}$ channel, where only two core-hole intermediate states are reached. Moreover, the intermediate states reached are more than the intermediate states  reached in the O$^{+}$+O$^{2+}$ channel.    The intermediate  states  in the O$^{2+}$+O$^{2+}$ channel are different for each of the final O$_{2}^{4+}$ molecular states that are reached following the third transition. Also, in the O$^{2+}$+O$^{2+}$ channel, there are significantly more final states than the 14 ones in the O$^+$+O$^+$ channel but even than the final states in O$^{+}$+O$^{2+}$.  Indeed, the final states of  the O$^{2+}$+O$^{2+}$ channel shown in  Fig.~\ref{fig:KER22} B  account for roughly 50\% of the probability for the O$^{2+}$+O$^{2+}$ channel to occur, while a similar number of final states in the O$^{+}$+O$^{2+}$ in  Fig.~\ref{fig:KERO++O2+} B and   Fig.~\ref{fig:KERonly} (a) contributes 74\%.  There are  several more final O$_{2}^{4+}$ states contributing less than 1\% each to the formation of O$^{2+}$+O$^{2+}$.

Comparing the peaks of the KER in Fig.~\ref{fig:KER22} A with the peaks of the KER of each final O$_{2}^{4+}$ molecular state in Fig.~\ref{fig:KER22} B, we  find that there is not   a clear correspondence except for a few peaks, i.e. the correspondence is worse than the one for the O$^{+}$+O$^{2+}$ channel.  Indeed, 
the KER spectra in Fig.~\ref{fig:KER22} A for all events leading to the O$^{2+}$+O$^{2+}$ channel have very wide peaks which are not well defined. Also, the KER spectra of each individual final O$_{2}^{4+}$ state in Fig.~\ref{fig:KER22} B 
are very wide. The peak at small energies of the KER, which was contributing significantly less than the higher energy peak  in   O$^{+}$+O$^{2+}$ (Fig.~\ref{fig:KERO++O2+} B), in   O$^{2+}$+O$^{2+}$, is almost as important as the peak at higher energies, with both peaks being wide.
  Nevertheless, there is still some correspondence between the total KER spectra and the KER spectra of the individual final states. Such is the case  for the 48.9 eV peak in  Fig.~\ref{fig:KER22} A and sub-plots (f) and (j) in Fig.~\ref{fig:KER22} B, and for the 43.4 eV peak and sub-plots (b), (e) and (n).

 \begin{widetext}
\begin{center}
   
  \includegraphics[scale=0.75]{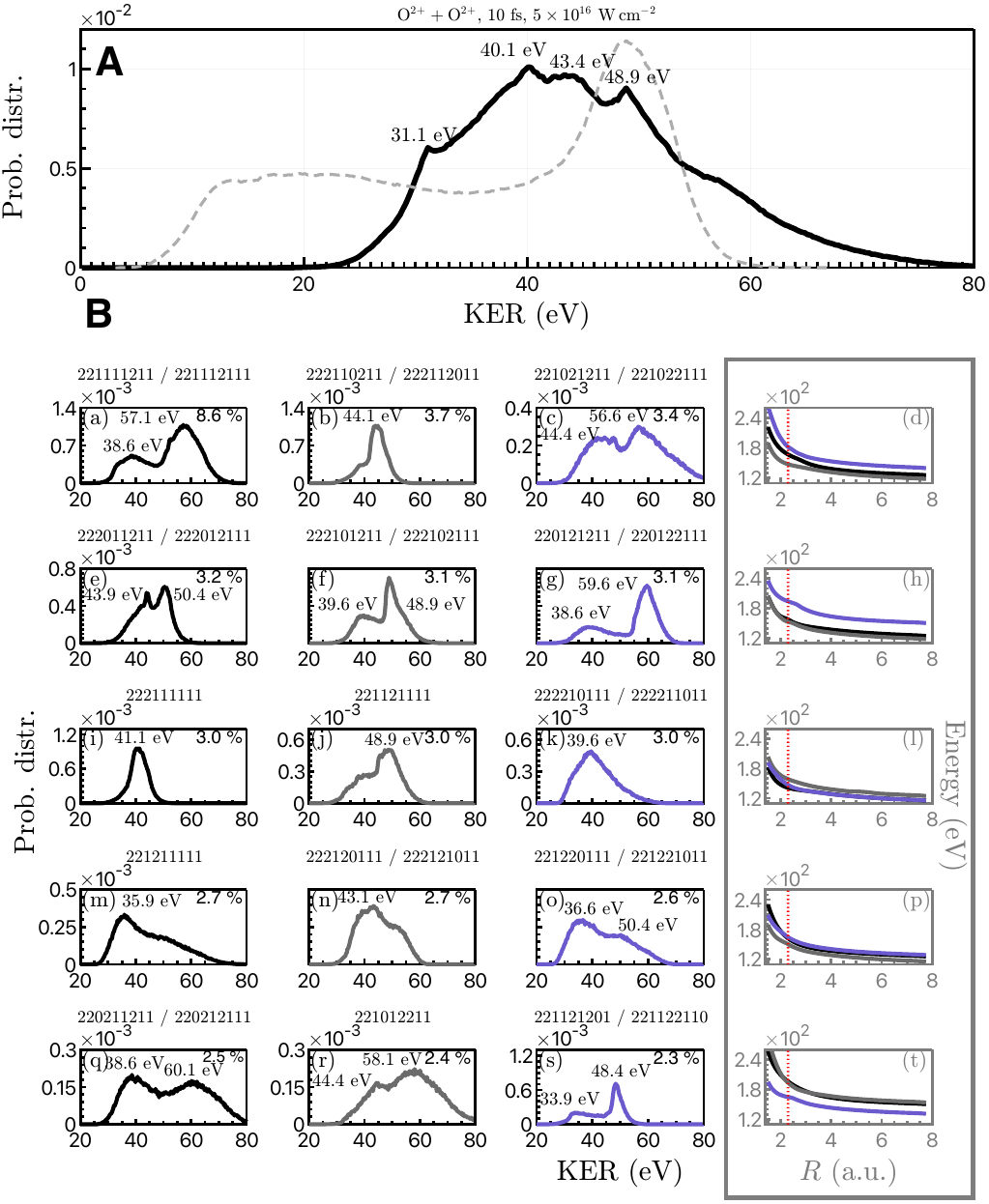}
    \captionof{figure}{As in Fig.~\ref{fig:figKERO+O+} A and Fig.~\ref{fig:figKERO+O+} B for the O$^{2+}$+O$^{2+}$ channel. The gray line shows the KER one would obtain if Coulomb explosion was to occur at the inter-nuclear distances the last A$_{CV}$ transition occurs. }
    \label{fig:KER22}
\end{center}
\end{widetext}

The peaks at larger values of the KER of the final states in Fig.~\ref{fig:KER22} B are obtained  by subtracting from the energy of the PEC of the final O$_{2}^{4+}$ state at 2.296 a.u. the dissociation energy limit.
 Next, we show that, as for the O$^{+}$+O$^{2+}$, we obtain the peaks at the higher energies of the spectra as follows
 For instance, we consider the peak at lower energies in the KER of the sub-plot (a) in  Fig.~\ref{fig:KER22} B.  We find that the distribution of the inter-nuclear distances, where the final electronic transition occurs, peaks at 3.45 a.u. for KER in the energy range from 20 eV to roughly 46 eV (not shown).  We then find that $\Delta E$, computed by subtracting from the energy of this final-state  PEC at 3.45 a.u. the asymptotic energy of the PEC, i.e. the dissociation energy limit, is equal to 29.65 eV. Also, the kinetic energy acquired 
at the inter-nuclear distance 3.45 a.u. is due to the molecule dissociating via the intermediate O$_{2}^{3+}$ molecular states that are reached after three electronic transitions. These transitions occur at inter-nuclear distances close to 2.295 a.u. Hence, to compute $\Delta E'$, we subtract from the energy of the intermediate-state PEC  at 2.296 a.u. the energy of the PEC  at 3.45 a.u. Since there are several intermediate molecular states, we  average the different  $\Delta E'$ using as weights the probability for each intermediate state to be reached after three electronic transitions.
 These intermediate states are shown in Table~\ref{tab:22a}.  
As a result, we obtain $\overline{\Delta E'} + \Delta E = 10.43+29.65=40.08 ~\mathrm{eV}$, which is close to the 38.6 eV peak  of the KER shown in sub-plot (a) in Fig.~\ref{fig:KER22} B. Following a similar procedure, we find that  $\overline{\Delta E'} + \Delta E = 12.73+27.04=39.77 ~\mathrm{eV}$ for the sub-plot (f) in Fig.~\ref{fig:KER22} B,  where $\overline{\Delta E'}$ is obtained in Table ~\ref{tab:22b}.

Hence, comparing the KER of O$^{+}$+O$^{+}$  in  Fig.~\ref{fig:figKERO+O+} A and B with the KER of O$^{+}$+O$^{2+}$ in Figs~\ref{fig:KERO++O2+} A and B and with the   KER of O$^{2+}$+O$^{2+}$  in Figs~\ref{fig:KER22} A and B, we find that  the channel with the higher charge of the atomic fragments has significantly  less well defined peaks and these peaks have less of a correspondence with the KER of the final molecular states. We also find that the KER of the channel with the higher charge states of the atomic fragments is more consistent with Coulomb explosion. Indeed, the KER of O$^{2+}$+O$^{2+}$  are closer to Coulomb-explosion KER than the KER of O$^{+}$+O$^{+}$, compare   gray lines and the actual total KER spectra in  Fig.~\ref{fig:figKERO+O+} B and Fig.~\ref{fig:KER22} B.
 The reason for the significantly less well defined KER spectra in O$^{2+}$+O$^{2+}$ is the large number of intermediate O$_{2}^{3+}$ molecular states that are reached after three electronic transitions as well as the large number of the final O$_{2}^{4+}$ molecular states that the molecule follows to dissociate. As we have seen, these intermediate states, which the molecule reaches just one electronic transition before the final one, as well as the final molecular states, which the molecule follows to dissociate, are increasing in number with increasing charge of the final atomic fragments.  
 Moreover, the final electronic transition occurs in the dominant pathways leading to a certain fragmentation channel at inter-nuclear distances that on average increase with increasing total charge of the  final atomic fragments. All the above lead to  KER spectra with wide and less well defined peaks for channels with higher total charge of the atomic fragments.





\begin{table}[H]
\centering
\footnotesize
\caption{
$\Delta E'$ corresponding to the  \texttt{221111211}/\texttt{221112111} states whose KER are shown in sub-plot (a) in Fig.~\ref{fig:KER22} B. We list the intermediate states reached by the dominant P$_{c}$P$_{c}$A$_{CV}$A$_{CV}$ pathway.
}
\label{tab:22a}
\begin{tabular}{ccc}
\multicolumn{3}{c}{} \\
\hline
Intermediate O$_{2}^{3+}$ states & Probability & $\Delta E'$ (eV) \\
\hline
\texttt{211122211} & 0.1423 & 10.83 \\
\texttt{121122211} & 0.1167 & 10.66 \\
\texttt{122112211} & 0.08768 & 13.07 \\
\texttt{212112211} & 0.07782 & 13.65 \\
\texttt{211212211} & 0.07728 & 11.76 \\
\texttt{122212111/122211211} & 0.07195 & 6.49 \\
\texttt{122122111/122121211} & 0.07185 & 10.97 \\
\texttt{212212111/212211211} & 0.06980 & 6.78 \\
\texttt{211222111/211221211} & 0.06746 & 6.53 \\
\texttt{121212211} & 0.06274 & 11.93 \\
\texttt{121222111/121221211} & 0.06056 & 6.72 \\
\texttt{212122111/212121211} & 0.06012 & 11.68 \\
\hline
& & $\overline{\Delta E'} = 10.43$ \\
\hline
\hline
\end{tabular}
\end{table}

\begin{table}[H]
\centering
\footnotesize
\caption{
$\Delta E'$ corresponding to the  \texttt{222101211}/\texttt{222102111} states whose KER are shown in sub-plot (f) in Fig.~\ref{fig:KER22} B. We list the intermediate states reached by the dominant P$_{c}$P$_{c}$A$_{CV}$A$_{CV}$ pathway.
}
\label{tab:22b}
\begin{tabular}{ccc}
\multicolumn{3}{c}{} \\
\hline
Intermediate O$_{2}^{3+}$ states & Probability & $\Delta E'$ (eV) \\
\hline
\texttt{212112211} & 0.1513 & 15.77 \\
\texttt{122112211} & 0.1441 & 15.17 \\
\texttt{122122111/122121211} & 0.1423 & 13.03 \\
\texttt{212122111/212121211} & 0.1352 & 13.77 \\
\texttt{122202211} & 0.1173 & 11.61 \\
\texttt{212202211} & 0.1101 & 11.83 \\
\texttt{212211211/212212111} & 0.0921 & 8.64 \\
\texttt{122212111/122211211} & 0.0894 & 8.32 \\
\hline
& & $\overline{\Delta E'} = 12.73$ \\
\hline
\hline
\end{tabular}
\end{table}

\section{Conclusions}
We study the interaction of the O$_{2}$ molecule with an XFEL pulse of 10 fs duration and  570 eV photon energy. This photon energy suffices to create molecular states with two core holes, one on  each core orbital. We compute the kinetic energy release (KER) of the three fragmentation channels O$^{+}$+O$^{+}$, O$^{+}$+O$^{2+}$ and O$^{2+}$+O$^{2+}$. We find that the KER of the low-charged channel has the sharpest peaks, with each peak corresponding to a peak in the KER of each final O$_{2}^{2+}$ molecular state, which dissociates to atomic fragments. 
Adding to each of these peaks the dissociation energy of the potential energy curve corresponding to the final state, one extracts the equilibrium inter-nuclear distance of O$_2$. The reason for the very good  correspondence between the peaks of the total KER and the KER of each final state is two-fold. Only two intermediate O$_{2}^+$  states are reached, which lead to  roughly 14 O$_{2}^{2+}$  molecular states that dissociate to two atomic fragments. Also,   
 we find that both the single photon ionization and  the Auger Meitner processes, taking place in the main pathway of O$^{+}$+O$^{+}$, occur around the equilibrium inter-nuclear distance of O$_{2}^{2+}$. 
 
 However, for the higher-charged fragmentation channels and particularly for O$^{2+}$+O$^{2+}$ the total KER are significantly wider with less well-defined peaks compared to the KER of     O$^{+}$+O$^{+}$. We find a poor correspondence between these wider peaks and the peaks of each of the final 
O$_{2}^{4+}$ molecular states that  dissociate to two atomic fragments. The reason is two fold. On one hand, the molecule transitions to a large number of intermediate O$_{2}^{3+}$ states, after three electronic transitions involving two single photon ionizations from the core orbitals  and an Auger-Meitner process, which finally lead to quite a large number of O$_{2}^{4+}$  molecular states. These multiple states blur the correspondence between the peaks in the total and individual  KER. This number of intermediate and final molecular states increases with increasing charge of the fragmentation channel. Finally, the inter-nuclear distances where the last electronic process takes place (for this channel an Auger-Meitner process) are on average larger than for lower-charged fragmentation channels. This results in more and wider peaks in the KER of each final molecular state and hence less well defined total KER spectra. 

The KER of the lower-charged fragmentation channels also significantly diverge from the KER one obtains due to Coulomb repulsion. Our findings are contrary to expectation from  Coulomb explosion imaging. In CEI  one expects that it is the KER of the higher charged fragmentation channels  that allow us to image best the initial equilibrium distance of the molecule. We find that it is the lower-charged ones that are not consistent with CEI that allow for best extraction of the equilibrium inter-nuclear distance. However, to achieve extraction of the inter-nuclear distance from the lower-charged fragmentation channels one needs to know the PECs of the molecule. We believe that our results demonstrate that some assumptions  in CEI should be reconsidered, such as that  rapid charge up of a molecule leads to fast fragmentation that allows imaging the initial inter-nuclear distance. Our results suggest that the last electronic transition during charge up of the molecule occurs at a wide range of inter-nuclear distances which leads to widening the KER. Moreover, our results show that higher charged channels involve significantly more molecular states which also result in less well-defined KER. We hope that these findings will initiate  new studies investigating the applicability of our findings to larger molecules.   

\section{Acknowledgments}
A.E gratefully acknowledges the Leverhulme Trust, grant number RPG-2025-179, for support of this work.

\bibliography{bibliography}{}

\end{document}